\documentclass[conference]{IEEEtran}
\IEEEoverridecommandlockouts
\usepackage{cite}
\usepackage{amsmath,amssymb,amsfonts}
\usepackage{algorithmic}
\usepackage{graphicx}
\usepackage{textcomp}
\usepackage{xcolor}

\usepackage{amsfonts,amsmath}
\usepackage{algorithm}

\usepackage{bm}
\usepackage{booktabs}
\usepackage{soul}
\usepackage{multirow}
\usepackage{subfig}
\usepackage[flushleft]{threeparttable}
\usepackage{adjustbox}
\usepackage{fancyhdr}
\usepackage{xcolor}
\usepackage[flushleft]{threeparttable}
\usepackage{tikz}
\usepackage{outlines}
\usepackage{balance}
\usepackage{tcolorbox}
\usepackage{wrapfig}
\usepackage{blindtext}
\usepackage{xcolor}
\usepackage{balance}
\def\BibTeX{{\rm B\kern-.05em{\sc i\kern-.025em b}\kern-.08em
    T\kern-.1667em\lower.7ex\hbox{E}\kern-.125emX}}
\begin{document}

\title{Accurate, Low-latency, Efficient SAR Automatic Target Recognition on FPGA

}

\author{
\IEEEauthorblockN{ Bingyi Zhang\IEEEauthorrefmark{1}, Rajgopal Kannan\IEEEauthorrefmark{2}, Viktor Prasanna\IEEEauthorrefmark{1}, Carl Busart\IEEEauthorrefmark{2}}
\IEEEauthorblockA{
    \IEEEauthorrefmark{1}University of Southern California \IEEEauthorrefmark{2}DEVCOM US Army Research Lab\\
    \IEEEauthorrefmark{1}\{bingyizh, prasanna\}@usc.edu \IEEEauthorrefmark{2}\{rajgopal.kannan.civ, carl.e.busart.civ\}@army.mil}
}

\maketitle

\begin{abstract}
Synthetic aperture radar (SAR) automatic target recognition (ATR) is the key technique for remote-sensing image recognition. The state-of-the-art convolutional neural networks (CNNs) for SAR ATR suffer from \emph{high computation cost} and \emph{large memory footprint}, making them unsuitable to be deployed on resource-limited platforms, such as small/micro satellites.

In this paper, we propose a comprehensive GNN-based model-architecture {co-design} on FPGA to address the above issues. \emph{Model design}:  we design a novel graph neural network (GNN) for SAR ATR.  The proposed GNN model incorporates GraphSAGE layer operators and attention mechanism, achieving comparable accuracy as the state-of-the-art work with near $1/100$ computation cost. Then, we propose a pruning approach including  weight pruning and input pruning. While weight pruning through lasso regression reduces most parameters without accuracy drop, input pruning eliminates most  input pixels with negligible accuracy drop. \emph{Architecture design}:  to fully unleash the computation parallelism within the proposed model, we develop a novel unified hardware architecture that can execute various computation kernels (feature aggregation, feature transformation, graph pooling). The proposed hardware design adopts the Scatter-Gather paradigm to efficiently handle the irregular computation {patterns} of various computation kernels. We deploy the proposed design on an embedded FPGA  (AMD Xilinx ZCU104) and evaluate the performance using MSTAR dataset. Compared with the state-of-the-art CNNs, the proposed GNN achieves comparable accuracy with $1/3258$ computation cost and $1/83$ model size. Compared  with the  state-of-the-art CPU/GPU, our FPGA accelerator achieves $14.8\times$/$2.5\times$ speedup (latency) and  is $62\times$/$39\times$ more energy efficient.

\end{abstract}


\section{Introduction}


Synthetic aperture radar (SAR) can acquire remote-sensing data  in all-weather conditions to observe target on the earth ground. SAR has been widely used in real-world applications, such {as} agriculture \cite{landuyt2018flood, zhan2021automated}, civilization \cite{li2021characterizing, zhang2020hyperli}, etc. SAR automatic target recognition (ATR) is the key technique to classify the target in a SAR image. Convolutional neural networks (CNNs) \cite{zhang2020convolutional, morgan2015deep, hu2018squeeze, pei2017sar, ying2020tai}  have been extensively studied  for ATR SAR since CNNs can extract discriminative features from an image. However, the CNN-based approaches \cite{zhang2020convolutional, morgan2015deep, hu2018squeeze, pei2017sar, ying2020tai}  suffer from two issues: (1) \emph{high computation cost}: to achieve high accuracy, the authors \cite{zhang2020convolutional, morgan2015deep, hu2018squeeze, pei2017sar, ying2020tai} develop large CNN models with high computation complexity, (2) \emph{large memory requirement}: these large CNN models have large number of parameters, which require large memory footprint.   Therefore, it is unsuitable to deploy large CNNs on resource-limited platforms, such as small/micro satellites \cite{bardi2014integration, septanto2019simulation, yokota2013newly, akbar2016parallel, tanaka2018development}.

The causes of the above issues are (1) heavy convolutional operations in CNNs, and (2) CNNs are hard to exploit data sparsity in SAR images because CNNs need to use the whole image as the input. In a SAR image (Figure \ref{fig:sar-image}), only {a} small set of pixels {belongs} to the target (defined as pixels of interest, POI), which can be easily extracted through applying a constant threshold \cite{zhu2020target}. However,  the extracted POI has irregular structure that {is} hard to be processed by CNNs, where Graph Neural Network (GNN) provides an opportunity. Intuitively, we can use the POI to construct a graph and use GNN to perform classification for the graph.
Fortunately, GNNs have been proven to be powerful models \cite{xu2018powerful} to classify graphs based on graph structural information and vertex features. Therefore GNNs
\cite{kipf2016semi, hamilton2017inductive, velivckovic2017graph} have been applied to many graph classification tasks  \cite{gligorijevic2021structure, zhu2020convsppis, zhao2021identifying, qi2017pointnet, qi2017pointnet++}. Recently, GNNs have been successfully applied to many image classification tasks \cite{ding2021semi, ding2021graph, nguyen2021modular}.
Motivated by that, we design a novel GNN model for SAR ATR (Section \ref{subsect:GNN-model-design}). We propose a graph representation $\mathcal{G}(\mathcal{V}, \mathcal{E})$ for a SAR image. The proposed GNN model can extract the structural information of the target from the constructed graph. To improve classification accuracy, we leverage the attention mechanism including spatial attention and channel attention to identify the important vertices and features. 
To further reduce the computation complexity, we  perform \emph{weight pruning} by training the GNN model through lasso regression and pruning the GNN model weights that have small absolute values.  Taking advantage of the GNN model, we perform \emph{input pruning} (POI extraction).  By eliminating the vertices that have small value, the computation complexity {is} reduced by $92.8\%$ with small accuracy loss ($<0.17\%$). 

The proposed GNN has the following advantages: (1) even without weight/input pruning, the proposed GNN has near $1/100$ computation cost as the state-of-the-art CNNs with similar accuracy, (2) while weight pruning can potentially be exploited by CNNs, input pruning is hard to be exploited by CNNs because CNNs need to use the whole image as the input. GNN is flexible to use a small set of input pixels as the input. Therefore, despite that we can accelerate the CNNs \cite{zhang2020convolutional, morgan2015deep, hu2018squeeze, pei2017sar, ying2020tai} on advanced CNN accelerators \cite{Xilinxdpu}, their latency is still significant (Section \ref{subsec:cmp-Latency}).

While the proposed GNN is lightweight that can be deployed on the resource limited platforms, accelerating GNNs is challenging.  GNNs have irregular computation pattern and heterogeneous computation kernels \cite{yan2020hygcn}, making them inefficient to be deployed on the general purpose processors. The pruned GNN model introduces additional irregularity through weight pruning. Moreover, the proposed model has various heterogeneous computation kernels (feature aggregation, feature transformation, graph pooling) that need to be mapped on an accelerator. While there are many GNN accelerators \cite{yan2020hygcn, zhang2021boostgcn, lin2022hp, zeng2020graphact, zhang2020hardware, zhang2021efficient, zhou2022model} proposed, none of them exploits the sparsity of the weight matrices or deals with graph pooling, which are still inefficient for the proposed model. While the proposed  GNN achieves high accuracy with small computation complexity, we believe that low-latency execution of SAR ATR must  be achieved through careful model-architecture co-design. 

 Therefore, we develop a novel unified hardware architecture for the proposed GNN model. We demonstrate the methods of mapping various computation kernels onto the proposed accelerator. In the accelerator design, we {adopt} Scatter-Gather paradigm to efficient deal with the irregular computation patterns of various kernels.  To the best of our knowledge, this is the first GNN-based model-architecture co-design for SAR ATR. 
Our main contributions are: 

\begin{itemize}
    \item We propose a lightweight GNN for SAR ATR that achieves comparable accuracy as state-of-the-art GNNs with  significant less computation complexity.
    \item We perform weight pruning and input pruning to dramatically reduce the computation complexity and the number of model weights.
    \item We design a unified hardware architecture that can execute various computation kernels in the proposed model. We {adopt} Scatter-Gather paradigm to deal with the irregular computation patterns.
    \item Taking advantage of the proposed hardware mapping strategy, we further optimize the load balance of various computation kernels (Section \ref{subsect:load-balance}). 
    \item We deploy our co-design on Xilinx ZCU104. We evaluate our co-design using MSTAR dataset. Compared with the state-of-the-art CNNs, the proposed GNN achieves comparable accuracy with $1/3258$ computation cost and $1/83$ model size. Compared  with the  state-of-the-art CPU/GPU, our FPGA accelerator achieves $14.8\times$/$2.5\times$ speedup (latency) and  is $62\times$/$39\times$ more energy efficient. 
\end{itemize}

\section{Background and Related Work}

\subsection{Related Work}

 \begin{figure}[ht]
    \centering
    \includegraphics[width=8cm]{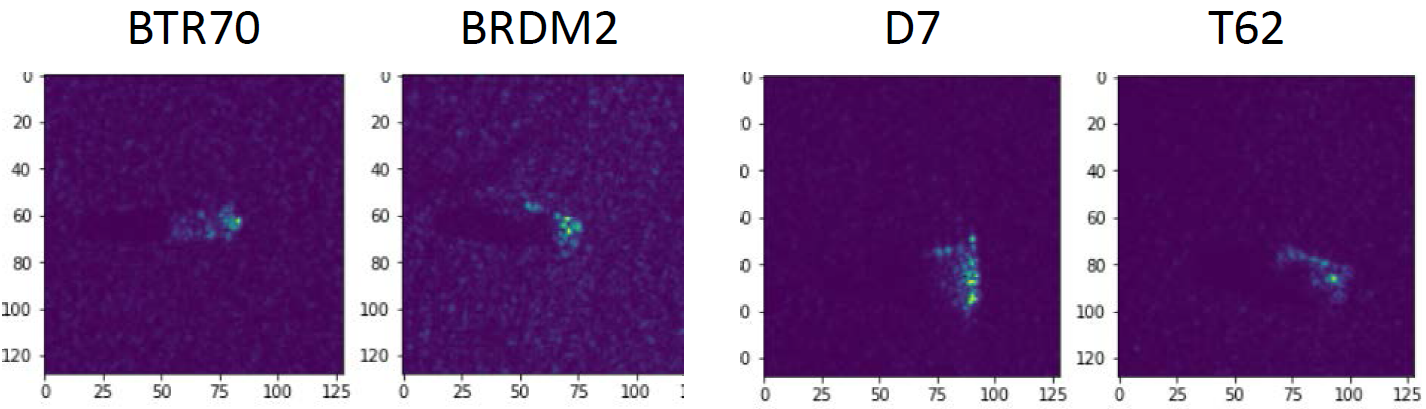}
    \caption{The SAR images of various targets (vehicles)}
     \label{fig:sar-image}
\end{figure}

SAR ATR is to automatically classify the target in a given SAR images (Figure \ref{fig:sar-image}). 
To achieve high accuracy, deep learning based methods have been extensively studied. David \cite{morgan2015deep} demonstrates that CNNs outperform  traditional methods, such as Support Vector Machine, etc. 
TAI-SARNET  \cite{ying2020tai} is a CNN model that incorporates  atrous convolution and inception module to achieve high accuracy for SAR ATR. The authors  \cite{pei2017sar} combine multi-view features to classify the target in SAR images. The authors \cite{zhang2020convolutional} propose the Convolutional Block Attention Module by exploiting the spatial attention and channel attention. However, the state-of-the-art CNNs \cite{ying2020tai, pei2017sar, zhang2020convolutional} suffer from high computation cost, making them unsuitable to be deployed on resource-limited platforms. Recently, the authors \cite{zhu2020target} {exploit} GNN for SAR ATR. They construct graphs from SAR images by connecting the {pixels} by the declined order of pixel grayscale value. However, the {constructed} graphs lose the structural information of targets, making it extremely sensitive to the variations of input pixel values.

\subsection{Graph Neural Network}

\begin{table}[]
\centering
\caption{Notations}
\vspace{-0.1cm}
\begin{adjustbox}{max width=0.48\textwidth}
\begin{tabular}{cc|cc}
\toprule
 \textbf{{Notation}} & \textbf{{Description}}  & \textbf{{Notation}}  & \textbf{{Description}} \\
 \midrule
\midrule
{$  \mathcal{G}(\mathcal{V},\mathcal{E},\bm{X^{0}})$ }& {input graph}  &  $ v_{i}$ & {$i^{th}$ vertex} \\ \midrule
$ \mathcal{V}$ &  {set of vertices} &  $ e_{ij}$ & {edge from $ v_{i}$ to $  v_{j}$} \\ \midrule
$ \mathcal{E}$& {set of edges} &  $ L$&{number of GNN layers} \\ \midrule
$ \bm{h}_{i}^{l}$& feature vector of $ v_{i}$
at layer $l$    &  $ \mathcal{N}(i)$& {neighbors of $ v_{i}$} \\ 

 \bottomrule
\end{tabular}
\end{adjustbox}
\vspace{-0.3cm}
\label{tab:notations}
\end{table}

The notations are defined in Table \ref{tab:notations}. Graph Neural Networks  (GNN) \cite{kipf2016semi, hamilton2017inductive, velivckovic2017graph} are proposed for representation learning on  graph $  \mathcal{G}(\mathcal{V},\mathcal{E},\bm{X}^{0})$. GNNs can learn from the structural information and vertex/edge features of the graph, and embed these information into low-dimension vector representation/graph embedding (For example, $\bm{h}^{L}_{i}$ is the embedding of vertex $v_{i}$). The vector representation can be used for many downstream tasks,  such as node classification \cite{hamilton2017inductive,kipf2016semi}, link prediction  \cite{zhang2018link}, graph classification \cite{ying2018hierarchical}, etc.  GNNs follow the  message-passing paradigm that vertices recursively aggregate information from the neighbors, for example:

\noindent \textbf{GraphSAGE}: GraphSAGE is proposed in \cite{hamilton2017inductive} for inductive representation learning on graphs. The GraphSAGE layer follows the aggregate-update paradigm:

\begin{small}
\begin{equation}
    \begin{split}
        & \text{aggregate:} \bm{z}_{i}^{l}  = \text{Mean} \left(  \bm{h}_{j}^{l-1}:j\in\mathcal{N}(i) \cup \{i\} \right) \\
        & \text{update:} \bm{h}_{i}^{l}  = \text{ReLU}  \left(\bm{z}_{i}^{l}\bm{W}_{\text{neighbor}}^{l} + \bm{b}_{\text{neighbor}}^{l} || \bm{h}_{i}^{l-1}\bm{W}_{\text{self}}^{l} + \bm{b}_{\text{self}}^{l} \right)
    \end{split}
\end{equation}
\end{small}

\section{Model-Architecture Co-Design}

To achieve accurate and efficient SAR ATR on FPGA platform, we perform comprehensive model-architecture co-design. The proposed co-design consists of a novel GNN model for SAR ATR (Section \ref{subsect:GNN-model-design}), a pruning strategy to reduce the computation complexity (Section \ref{subsect:network-pruning}), a novel hardware design to efficiently execute the proposed GNN (Section \ref{subsec:architecture-design}), and the strategy to keep load balance within various computation kernels (Section \ref{subsect:load-balance}).  The key novelty of our hardware design is that it can execute various computation kernels in the proposed model, and it can efficiently handle the irregular computation patterns caused by the sparsity of weight matrices. 
We use the widely used MSTAR dataset \cite{master} for performance evaluation. We target various performance metrics: (1) \emph{Accuracy}: the accuracy on MSTAR dataset, (2) \emph{Computation complexity}: the total computation complexity for inferring a SAR image, (3) \emph{Number of parameters}: the total number of parameters in the model, (4) \emph{Latency}: the latency for inferring a SAR image, (5) \emph{Energy Consumption}: the energy consumption for inferring a SAR image.

\subsection{GNN Model Design}
\label{subsect:GNN-model-design}

 \begin{figure}[h]
    \centering
    \includegraphics[width=7cm]{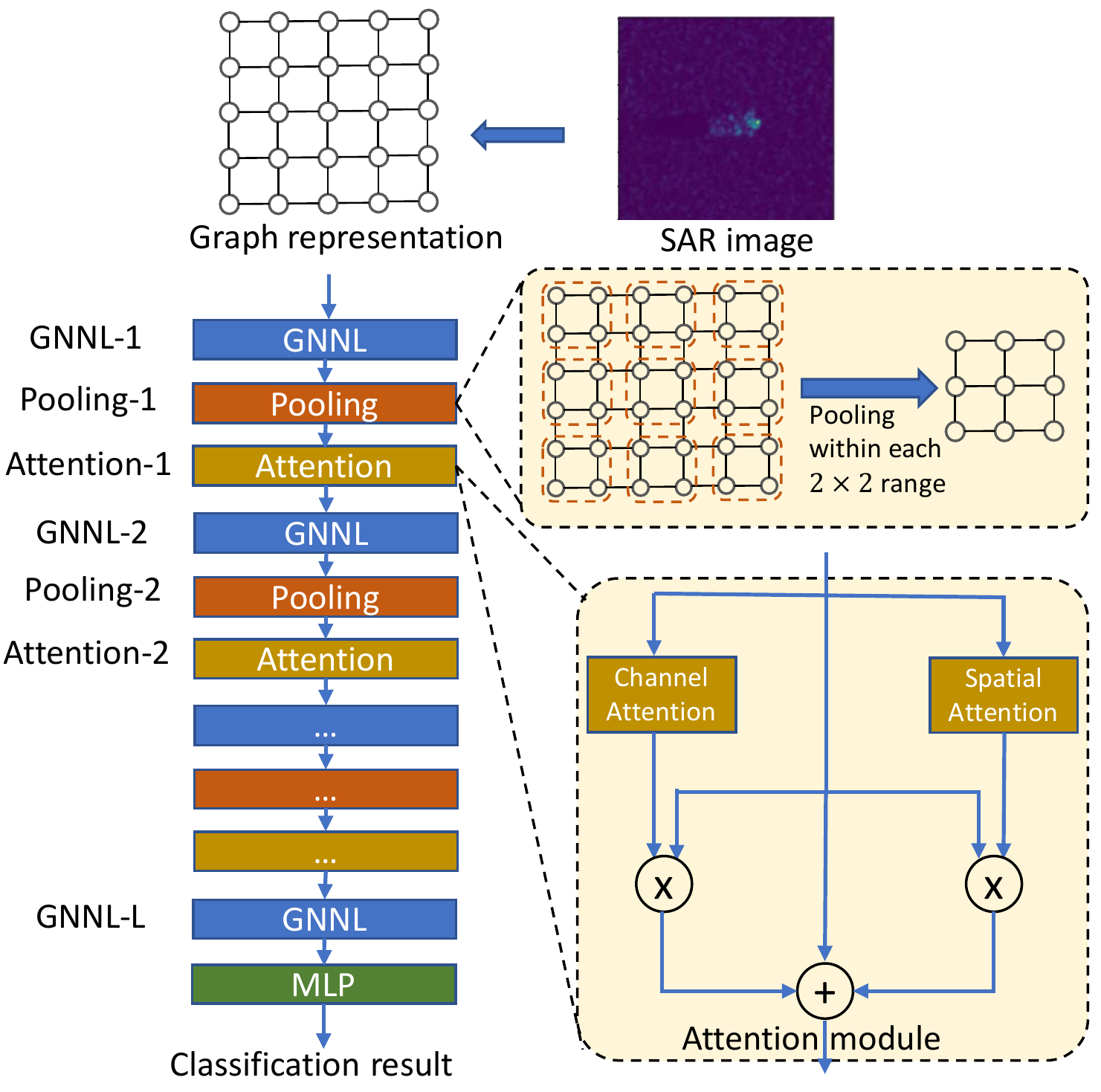}
    \vspace{-0.2cm}
    \caption{ The proposed GNN model}
    \vspace{-0.3cm}
     \label{fig:gnn-structure}
\end{figure}

\noindent \textbf{Graph representation}: We represent a SAR image as a graph $\mathcal{G}(\mathcal{V}, \mathcal{E})$, with each pixel viewed as a vertex. Each pixel/vertex is connected to its four neighbors (up, down, left, right) with edges. The feature of a vertex is the grayscale value of the pixel. Such graph representation maintains structural information of the target that can be learned by GNN for classification. It also provides the opportunity for input pruning (Section  \ref{subsect:network-pruning}).

\vspace{0.1cm}
\noindent \textbf{GNN model}: As shown in Figure \ref{fig:gnn-structure}, the proposed GNN model has a sequence of layers, including GNN layer (GNNL), graph pooling layer (Pooling), Attention module (Attention). For GNN layer, we use the GraphSAGE layer operators \cite{hamilton2017inductive}, which {have been} proven to achieve superior accuracy in various application domains.  For graph pooling layer, since the input graph has 2-D grid structure, we adopt the similar pooling strategy as the CNN for 2-D image. Within each local $s\times s$ range having $s^{2}$ vertices, the pooling operator (e.g., Max(), Min()) is performed on the $s^{2}$ vertices to obtain an output vertex.  Figure \ref{fig:gnn-structure} demonstrates the pooling operation of size $2\times 2$ with stride $2$. Motivated by the attention mechanism in CNN \cite{woo2018cbam},  the proposed Attention module consists of a  Channel Attention module and a Spatial Attention module. Suppose the input to Attention Module is $\{\bm{h}_{i}:v_{i}\in \mathcal{G}\}$, where $\bm{h}_{i}\in \mathbb{R}^{c}$ is the feature vector of $v_{i}$ and $c$ is the length of the feature vector. The Channel Attention calculates the attention score $\bm{F}_{ch}$ of each feature through a Multi-layer perceptron. 
Then, each vertex is multiplied by $\bm{F}_{ch}$ to obtain $\{(\bm{h}_{i})':(\bm{h}_{i})' = \bm{h}_{i} \otimes \bm{F}_{ch},  v_{i}\in \mathcal{G}\}$ where $\otimes$ is the element-wise multiplication. The Spatial Attention module calculates the attention score of each vertex using a GNN layer (GraphSAGE layer operators):
$$
    \{\alpha_{i}:v_{i}\in \mathcal{G}\}= \text{sigmoid}(\text{GNNL}(\{\bm{h}_{i}:v_{i}\in \mathcal{G}\})), 
$$
Then, each vertex feature vector is multiplied by its attention score: $\{(\bm{h}_{i})'':(\bm{h}_{i})'' = \alpha_{i}\bm{h}_{i},  v_{i}\in \mathcal{G}\}$. The output of the Attention module is calculated by:
\begin{equation}
    \{\bm{h}_{i}^{\text{output}}:\bm{h}_{i}^{\text{output}} = \bm{h}_{i} + (\bm{h}_{i})' + (\bm{h}_{i})'', v_{i}\in \mathcal{G}\}
\end{equation}
After GNNL-$L$, all the feature vectors are flattened to a vector which becomes the input to the last MLP (Multi-layer Perceptron) for classification.

\subsection{Network Pruning}
\label{subsect:network-pruning}

\noindent \textbf{Weight pruning}: To reduce the total computation complexity,  we perform weight pruning by training the model using lasso regression \cite{tibshirani1996regression}. We add a L1 penalty term to the loss function:
$$
  \text{loss} = \sum_{i=1}^{N}(y_{i} - \text{Model}(\mathcal{G}_{i}))^{2} + \lambda \sum_{w}^{W}|w|
$$
The penalty term results in weight shrinkage. Some model weights  become zeros and are eliminated from the model.
After training, we set a threshold $I_{\text{weight}}$ and the weights with absolute {values} smaller than $I_{\text{weight}}$ are pruned.

\vspace{0.1cm}
\noindent \textbf{Input pruning}: In a SAR {image}, most pixels outside of the target have negligible grayscale values. Therefore, in the graph representation $\mathcal{G}(\mathcal{V}, \mathcal{E})$ of a SAR image, we set a threshold $I_{\text{vertex}}$ and  prune the vertices that {have} grayscale values smaller than $I_{\text{vertex}}$. The edges connected to the pruned vertices are also pruned. After input pruning, the eliminated vertices maintain the same positions in the graph pooling layer and do not participate in the pooling operation. For example, in a local $2\times 2$ range, if a vertex is pruned, the pooling operator will  operate on the {remaining} three vertices. For  the input to last MLP, the feature vectors of the pruned vertices are padded using zeros.

\subsection{Architecture design}
\label{subsec:architecture-design}

 \begin{figure*}[ht]
    \centering
    \includegraphics[width=18cm]{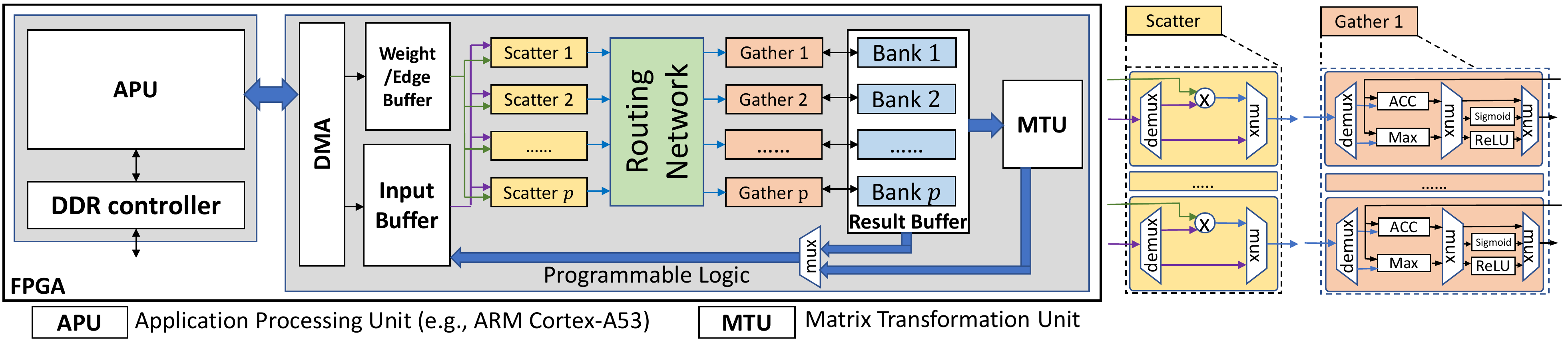}
    \vspace{-0.2cm}
    \caption{ The diagram of the system architecture}
    \vspace{-0.3cm}
     \label{fig:hardware-architecture}
\end{figure*}

The objective of the architecture design is to (1) support various computation kernels in the proposed model, (2) handle the irregular computation patterns caused by the feature aggregation in the GNN layer and the sparsity of the weight matrices. 
Figure \ref{fig:hardware-architecture} shows the proposed architecture design on the embedded FPGA platform. The system consists of an Application Processing Unit (APU) and an FPGA accelerator in Programmable Logic Region. The FPGA accelerator executes the inference process of the GNN model. In the FPGA accelerator, there is a Weight/Edge Buffer (WEB) to store the model weights and edges of input graph, an Input Buffer (IB) to store the input vertex feature vectors, a Results Buffer (RB) to store the  output vertex feature vectors. The Matrix Transformation Unit (MTU) performs matrix transformation to prepare the require data layout for the next layer. 
Thanks to the proposed {lightweight} model, the trained model is fully stored in the Weight Buffer, eliminating the memory traffic of loading the model weights at runtime. 

\noindent \textbf{Run Time}: At runtime, the APU receives an input SAR image and transform it into the graph presentation. During the transformation, the pixels that have grayscale value smaller than $I_{\text{vertex}}$ are pruned. Then, the APU sends the input graph to the Input Buffer of the accelerator.  The accelerator executes each layer using Scatter-Gather paradigm (SGP). The accelerator exploits the computation parallelism within each layer. After finishing the execution of all layers, the accelerator sends the classification result back to the APU.

\section{Hardware Mapping}
\subsection{Computation kernels}
We categorize the computation kernels  into two classes: 

\noindent \textbf{Vertex aggregation kernel (VAK)}:  VAKs include (1) feature aggregation (in GNN layer, and in Spatial Attention module) (2) graph pooling.  In VAKs, each vertex propagates its feature vector to the neighbors or within a local range (graph pooling). 

\noindent \textbf{Vertex updating kernel (VUK)}:  VUKs include  (1) feature update (in GNN layer, and in Spatial Attention module) (2) Channel attention of Attention module,  (3) the last MLP. In the VUKs, the feature vector of each vertex is multiplied by a weight matrix to obtain the updated feature vector.  Due to our weight pruning, the weight matrices have high data sparsity (1\%-33\% data density).

\subsection{Kernel Mapping using Scatter-Gather Paradigm}

\begin{algorithm}
\caption{Scatter-Gather paradigm}\label{alg:scatter-gather}
\begin{small}
\begin{algorithmic}
\WHILE{not done}
    \STATE \emph{Scatter Unit}:
    \FOR{each edge $e\langle src,dst,weight \rangle$ }  
        \STATE Produce update $u \gets ${Scatter($src.vector, e.weight$)}
    \ENDFOR
    \STATE \emph{Gather Unit}:
    \FOR{each update $u\langle dst,vector \rangle$ }
        \STATE Update vertex $v_{dst} \gets$ {Gather($u.vector$)}
    \ENDFOR
    \ENDWHILE
\end{algorithmic}
\end{small}
\end{algorithm}

 \begin{figure}[h]
    \centering
    \includegraphics[width=7.5cm]{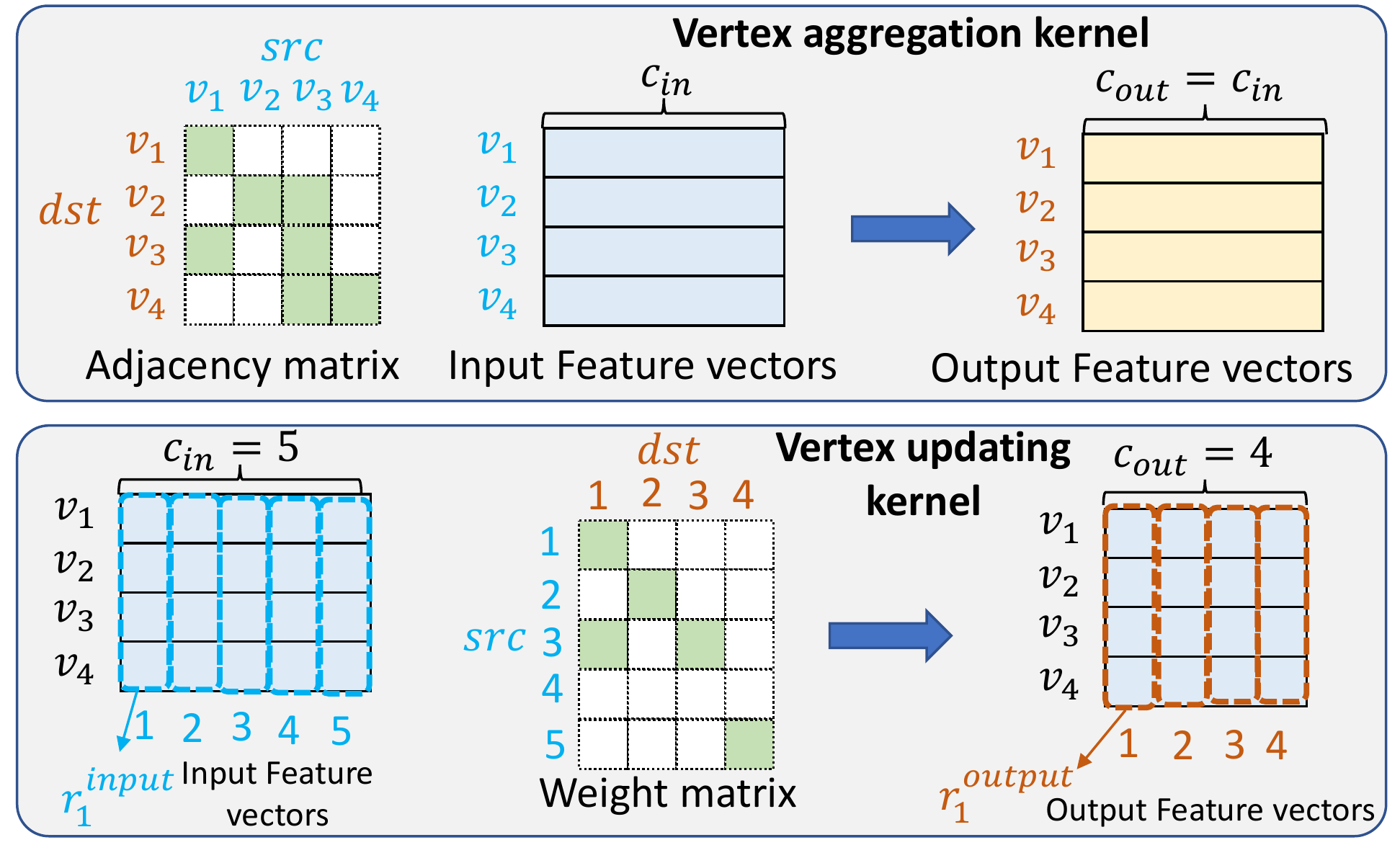}
    \vspace{-0.2cm}
    \caption{ The diagram of mapping the two types of kernels using Scatter-Gather paradigm}
    \vspace{-0.3cm}
     \label{fig:diagram-of-two-kernels}
\end{figure}

The accelerator design is based on the Scatter-Gather paradigm (Algorithm \ref{alg:scatter-gather}). There are $p$ parallel pipelines. Each pipeline consists of a Scatter Unit  and  a Gather Unit. The Routing Network routes the intermediate results to the destination based on index $dst$. 
To map the VAKs and VUKs to the accelerator, we propose the following mapping strategy (An example is shown in Figure \ref{fig:diagram-of-two-kernels}):

\noindent \textbf{Mapping VAK}: VAK can be directly mapped to the accelerator. For each edge
$e\langle src,dst,weight \rangle$, the Scatter Unit loads the feature vector of $v_{src}$ from input buffer and produces an update $u\langle dst,vector \rangle$. The update $u\langle dst,vector \rangle$ is routed to the corresponding Gather Unit and the Gather Unit applies the update to the destination vertex $v_{dst}$.

\noindent \textbf{Mapping VUK}: For VUK, we group a batch of vertices $batch$ and the feature vector of each vertex $\{\bm{h}_{i}^{\text{input}}:v_{i}\in batch\}$ is multiplied {by} the weight matrix $\bm{W}$ simultaneously.  The output feature vectors are $\{\bm{h}_{i}^{\text{output}}:\bm{h}_{i}^{\text{input}}\bm{W}, v_{i}\in batch\}$. To apply the Scatter-Gather paradigm, we perform feature concatenation.  For example, we concatenate the first feature of each vertex $\{\bm{h}_{i}(1):v_{i}\in batch\}$ as a vector $\bm{r}_{1}^{input}$. The vector $\bm{r}_{1}^{input}$ has $src$ index $1$ since its contains the $1^{st}$ feature of each input feature vector. For the weight matrix $\bm{W}$, we represent each non-zero element in the weight matrix as an edge $e\langle src,dst,weight \rangle$. During execution, for each non-zero weight  $e\langle src,dst,weight \rangle$, the Scatter Unit loads the $\bm{r}_{src}^{input}$ from the input buffer and produces an update $u\langle dst,vector = e.weight \times \bm{r}_{src}^{input}\rangle$. Then, the Gather Unit applies the update $u\langle dst,vector\rangle$ to the destination $\bm{r}_{dst}^{output}$. $\bm{r}_{dst}^{output}$ contains the $dst^{th}$ features of each output feature vector in the $batch$.

Note that VAK and VUK have different data {layouts}.  In VAK, the input/output feature vectors are stored in vertex-major order. In VUK, the {input}/output feature {vectors} are stored in feature-major order. To switch between the two {data}
layouts, we implement a Matrix Transformation Unit (MTU) to perform data layout transformation.

\subsection{hardware modules}
\noindent \textbf{Scatter/Gather Unit}: A Scatter Unit has an array of $q$ processing elements. Each processing element has a multiplier to perform the multiplication between an edge/weight and a vertex feature. Similar to the Scatter Unit, a Gather Unit has an array of $q$ processing elements. Each processing element has an Accumulator (ACC), a Max Unit, a ReLU Unit, a sigmoid Unit.  The multiplexer (MUX) and demultiplexer (DEMUX)  select the datapath for the current layer.

\noindent \textbf{Routing Network}: The routing network is implemented using a hardware-efficient butterfly network \cite{choi2021hbm}.

\noindent \textbf{Sigmoid Unit}: We exploit the piecewise linear approximation (PLA) \cite{zhang2017implementation} for Sigmoid Function. 


\section{Load Balance and Performance Model}

\subsection{Load Balance}
\label{subsect:load-balance}
\noindent \textbf{Load balance in VAK}: The workload balance of VAK depends on how to partition the vertices into $p$ memory banks of the Result Buffer. Load imbalance is a significant issue in GNN \cite{geng2020awb} if the graph has highly imbalanced degree distribution. Thanks to our graph {representation}, the vertices in the graph have degrees ranging from $0$ to $4$. We use a greedy approach to keep the load balance of the $p$ parallel {pipelines}. 
For VAK, the destination vertices that have  same degree $i~(0\leqslant i \leqslant 4)$ are evenly partitioned into $p$ banks of the Result Buffer. Through the proposed partitioning strategy, each pipeline has the same amount of workload. The graph partitioning has a small overhead $\mathcal{O}(|\mathcal{V}|L_{p})$ and is performed by the APU, where $L_{p}$ is the number of graph pooling layers in the model. The proposed partitioning algorithm can be easily parallelized using multiple threads on APU.

\noindent \textbf{Load balance in VUK}: {To} execute VUK, we need to partition the weight matrix along the $dst$ dimension (Figure \ref{fig:diagram-of-two-kernels}). Each Gather Unit is responsible for accumulating the partial results of a partition. To achieve perfect load balance, each partition should have the same number of non-zero elements. Since  the partitioning of weight matrix is an offline process, we are able to adopt complexity algorithm to find the near optimal data partitioning. In this work, we exploit Longest-processing-time (LPT) first algorithm that is proved to achieve $4/3$ approximation factor \cite{eck1993minimization} to the optimal partition solution.

\subsection{Performance Model}

\noindent \textbf{Modeling VAK}: For a VAK kernel, the length of input feature vector $c_{\text{in}}$ is same {as} the length of output feature vector $c_{\text{out}}$: $c_{\text{in}} = c_{\text{out}}$. A Scatter Unit or a Gather Unit can process $q$ features in each clock cycle. The $p$ parallel pipelines can process $p$ edges simultaneously. Therefore, the execution time of a VAK kernel is:
\begin{equation}
    t_{\text{VAK}} = \left\lceil \frac{|\mathcal{E}|}{p} \right\rceil \cdot  \left\lceil \frac{c_{\text{in}}}{q} \right\rceil
\end{equation}

\noindent \textbf{Modeling VUK}: 
To execute a VUK, the accelerator groups a batch of $q$ vertices at a time to fully {utilize} the Scatter Unit/Gather Unit. The $p$ parallel  {pipelines} can process $p$ non-zero elements in the weight matrix. Therefore, the execution time of a VUK kernel is:
\begin{equation}
    t_{\text{VUK}} = \left\lceil \frac{|\mathcal{V}|}{q} \right\rceil  \cdot \left\lceil \frac{\text{nnz}(\bm{W})}{p} \right\rceil  
\end{equation} 
where $\text{nnz}(\bm{W})$ is the number of non-zero elements in the weight matrix $\bm{W}$. 
Since our accelerator exploits the computation parallelism within each kernel, the total execution time is the sum of the execution time of all kernels and preprocessing overhead.

\section{Implementation and Experimental Results}

\subsection{Implementation Details and Resource Utilizations}
We implement our accelerator on an embedded FPGA platform -- Xilinx ZCU104. We implement 8 pipelines (8 Scatter Units and 8 Gather Units). Each Scatter/Gather Unit has 16 processing elements (PEs). In a Scatter Unit, a PE consumes 3 DSPs and in a Gather Unit, a PE consumes 7 DSPs.  The routing network has 8 input ports and 8 output ports. Each port is 512-bit that can  { receive/send 16 32-bit data}. The APU is a quad-core ARM-A53 processor running at 1.3 GHz. The accelerator is developed using High-Level Synthesis. The accelerator consumes 1280 DSPs, 96 URAMs, 221 BRAMs,  178K LUTs. The accelerator runs at 125 MHz. The resource utilization and frequency are reported after Place\&Route.

\subsection{Benchmark and Baseline Platform}
\noindent \textbf{Benchmark}:
We conduct experiments using the widely used MSTAR dataset. The setting of MSTAR dataset follows the state-of-the-art {work} \cite{zhang2020convolutional, pei2017sar, ying2020tai, morgan2015deep}. The dataset contains the SAR images of 10 classes of ground vehicles. The training set has 2747 images and the testing set has 2427 images. Each SAR image has size $128\times128$ and each pixel has {a} grayscale value indicating the magnitude of the SAR signal.
\begin{table}[!ht]
\centering
\vspace{-0.3cm}
\caption{Specifications of various platforms }
\vspace{-0.1cm}
\begin{threeparttable}

\begin{adjustbox}{max width=0.47\textwidth}
\begin{tabular}{c|ccc}
 \toprule
\textbf{Platforms} & \begin{tabular}[|c|]{@{}c@{}} CPU \\  AMD Ryzen 3990x \end{tabular}  & \begin{tabular}[|c|]{@{}c@{}} GPU \\  Nvidia RTX3090 \end{tabular} & \begin{tabular}[|c|]{@{}c@{}} FPGA \\   ZCU 104 \end{tabular}  \\ 
\midrule \midrule 

 {Release Year}    &  2020 &  2020 & 2018   \\
 {Technology}  & TSMC 7 nm   & TSMC 7 nm & TSMC 16 nm  \\ 
{Frequency} & 2.9 GHz  & 1.7 GHz & 125 MHz 
      \\ 
{On-chip Memory}& 256 MB L3 cache & 6 MB L2 cache & 4.8 MB  \\
 \bottomrule
\end{tabular}
\end{adjustbox}
\end{threeparttable}
\label{tab:platform-specifications}
\vspace{-0.3cm}
\end{table}

\noindent \textbf{Baseline Platform}: We compare our performance with the state-of-the-art CPU and GPU platforms as shown in Table \ref{tab:platform-specifications}. On the CPU platform and GPU platform, we run the proposed model using Pytorch Geometry (PyG) \cite{Fey/Lenssen/2019} of 1.8.0 version. For CPU platform, PyG uses the Intel MKL as the backend and for the GPU platform, PyG uses the CUDA 11.1 as the backend. To exploit the sparsity of the weight matrices on the CPU and GPU platforms, we modify the GraphSAGE layer\footnote{https://pytorch-geometric.readthedocs.io/en/latest/ \\
\_modules/torch\_geometric/nn/conv/sage\_conv.html\#SAGEConv} of PyG 
by using the $\verb|torch.sspaddmm()|$ for efficient multiplication of feature vectors and sparse weight matrices.

\subsection{Accuracy, Computation Complexity, Model Size}

\noindent \textbf{Weight/Input pruning}: The magnitude of the SAR signal ranges from $0$ to $8$. we set the $I_{\text{vertex}}$ as $0.1$ {because} it can filter out most irrelevant  pixels.
\begin{table}[!ht]
\centering
\begin{adjustbox}{max width=0.47\textwidth}
\begin{tabular}{cccccc}
\toprule
 & Type & Accuracy & \# of FLOPs & \# of Para. & Model Size \\ \midrule \midrule
 \cite{zhang2020convolutional}&CNN&92.3\%&$\frac{1}{12}\times$&$0.5 \times 10^{6}$& 16 Mb \\ \midrule
\cite{pei2017sar} &CNN&97.97\%&$\frac{1}{10}\times$&$0.65 \times 10^{6}$  &  20.8 Mb \\  \midrule
\cite{ying2020tai}&CNN&98.52\%&$\frac{1}{3}\times$&$2.1 \times 10^{6}$  & 67.2 Mb \\  \midrule
\cite{morgan2015deep}&CNN&99.3\%&$1\times$ (6.94 GFLOPs)&$2.5 \times 10^{6}$  & 80 Mb \\ \midrule
 This work&GNN&99.09\%&$\frac{1}{3258}\times$&$0.03\times 10^{6}$  & 0.96 Mb\\ \bottomrule
\end{tabular}
\end{adjustbox}
\end{table}
We compare \emph{Accuracy}, \emph{computation complexity}, \emph{number of parameters} with state-of-the-art work  \cite{zhang2020convolutional, pei2017sar, ying2020tai, morgan2015deep}. Compared with the state-of-the-art CNN \cite{morgan2015deep}, the proposed model achieves comparable accuracy with only $\frac{1}{3258}$ computation complexity and $\frac{1}{83}$ number of parameters on average.

\subsection{Evaluation of Latency}
\label{subsec:cmp-Latency}

 \begin{figure}[h]
    \centering
    \vspace{-0.3cm}
    \includegraphics[width=7.5cm]{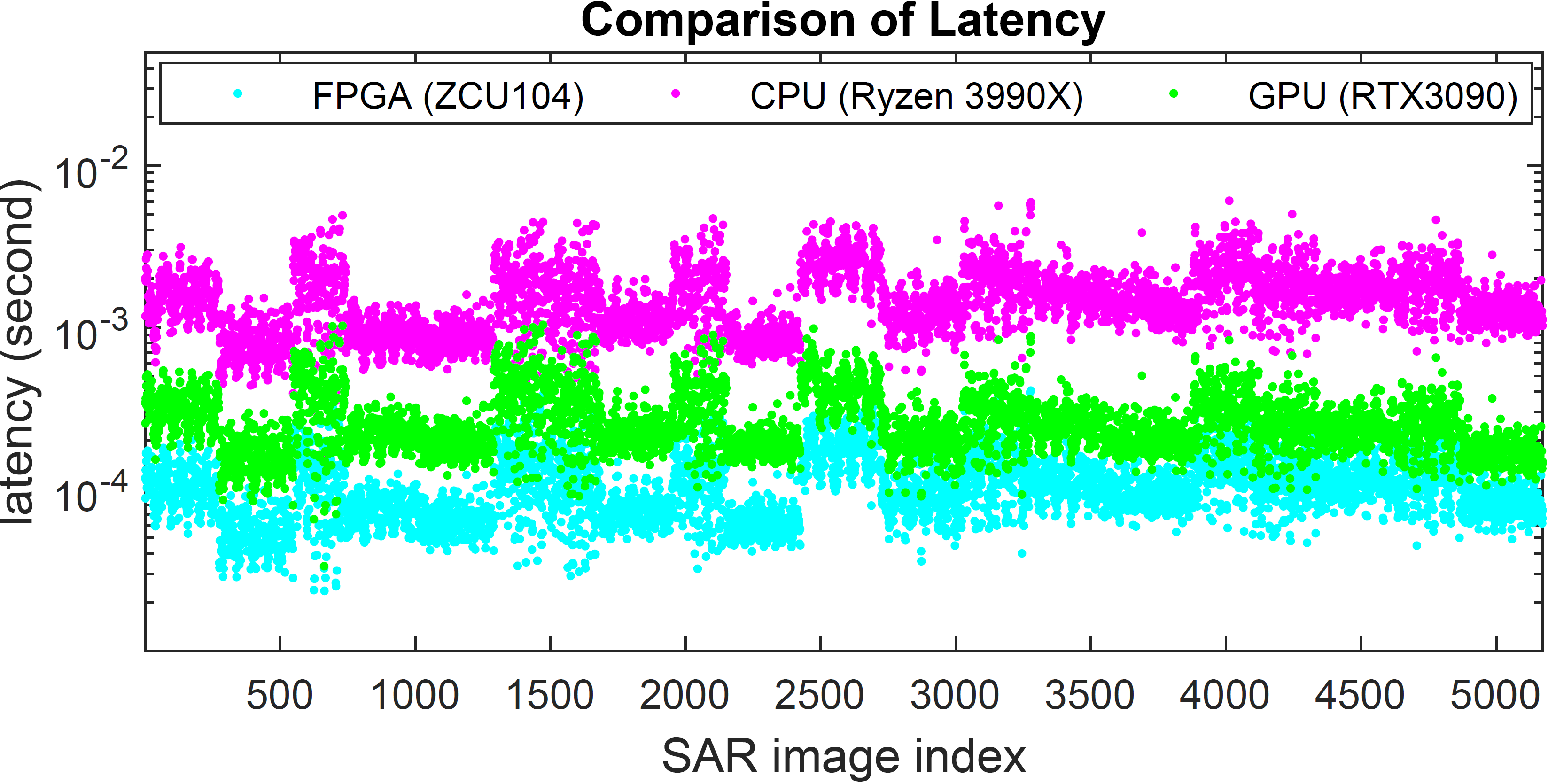}
    \vspace{-0.2cm}
    \caption{ X-axis is the index of the SAR image (training set + testing set). Y-axis is the inference latency of a SAR image.}
    \vspace{-0.3cm}
     \label{fig:latency-comparison}
\end{figure}


To compare the latency of various platforms, we set the batch size as 1. The measured latency on FPGA accelerator is end-to-end from the time when APU receives the SAR image to the time {when} APU gets the classification results from the accelerator, which means the preprocessing overhead is included in the measured latency. We measure the inference latency on all images in training and testing sets. The comparison results are shown in Figure \ref{fig:latency-comparison}. On average, our FPGA accelerator is $14.8\times$, $2.5\times$ faster than the CPU and GPU platforms in terms of latency. Since we use the input pruning, the graph representations of the images after input pruning have various number of vertices. Therefore, the inference latency fluctuates with images.
Compared with CPU/GPU, our accelerator has lower latency.  Because CPU/GPU has complex cache hierarchy and large cache latency (e.g., CPU has high cache latency: L3 cache 32ns, L2 cache 12ns). Therefore, loading feature vectors and weight matrices leads to large latency. In contrast, our FPGA accelerator can access data in one-clock cycle due to our customized on-chip memory organization. Moreover, our FPGA accelerator adopts the Scatter-Gather paradigm  to efficiently deal with irregular computation  in various computation kernels. 


\begin{table}[!ht]
\centering
\caption{Latency comparison on ZCU 104 and GPU }
\begin{adjustbox}{max width=0.47\textwidth}
\begin{tabular}{c|cccc|c}
\toprule
\textbf{Model} & \cite{zhang2020convolutional} & \cite{pei2017sar} & \cite{ying2020tai} &  \cite{morgan2015deep}& Proposed model\\ \midrule \midrule 
 & \multicolumn{4}{c|}{{[Xilinx DPU]}} & {{[Proposed design]}} \\  
ZCU104 & 0.88 ms & 1.23 ms & 3.09 ms & 12.1 ms  & 0.105 ms \\   \midrule \midrule  
GPU (RTX3090)& 1.53 ms & 2.5 ms & 9.5 ms & 31.2 ms  & 0.269 ms \\   \bottomrule 
\end{tabular}
\end{adjustbox}
\vspace{-0.3cm}
\label{tab:cmp-latency-soda}
\end{table}

\noindent \textbf{Impact of model design}: To compare the inference latency with the state-of-the-art CNNs, we deploy AMD Xilinx DPU \cite{Xilinxdpu} (2 * B4096 @ 300 MHz configuration) on the same FPGA platform (ZCU 104) to execute the CNN models in  \cite{zhang2020convolutional, pei2017sar, ying2020tai, morgan2015deep}. AMD Xilinx DPU is the state-of-the-art  FPGA overlay accelerator for CNNs. The average inference latency is shown in Table \ref{tab:cmp-latency-soda}. {The proposed GNN on the proposed design (The column 6 of Table \ref{tab:cmp-latency-soda}) is $115\times$  faster than \cite{morgan2015deep} on DPU}. Note that DPU uses 8-bit data quantization for the weights and activations. Our work uses 32-bit floating point data format. DPU has more computation parallelism by operating on 8-bit data. 



\noindent \textbf{Preprocessing Overhead}: {We} measure the preprocessing overhead on APU. For a SAR image, APU transforms it into graph representation (Section \ref{subsect:GNN-model-design}) with input pruning (Section \ref{subsect:network-pruning}), and graph partitioning (\ref{subsect:load-balance}). The average preprocessing time is 11.8 us for a SAR image, which is negligible compared with the total latency.

\begin{table}[!ht]
\centering
\caption{Comparison of Energy Consumption}
\vspace{-0.2cm}
\begin{adjustbox}{max width=0.4\textwidth}
\begin{tabular}{cccc}
\toprule
\textbf{Platform} & \textbf{Inference Speed} & \textbf{Power} & \textbf{Energy (mJ/image)} \\ \midrule 
Ryzen 3990X &  $644$ (image/s) & $26.5 W$ &  41.1 (mJ/image)  \\   \midrule 
Nvidia RTX3090 & $3717$ (image/s) & $97W$ & 26.0 (mJ/image) \\ \midrule 
ZCU104 & $9500$ (image/s) & $6.3 W$ & 0.66 (mJ/image)  \\ \bottomrule
\end{tabular}
\end{adjustbox}
\vspace{-0.3cm}
\label{tab:energy-comparison}
\end{table}

\subsection{Evaluation of Energy Consumption}

Table \ref{tab:energy-comparison} shows the comparison of energy consumption on various platforms. On the CPU platform, we measure the power consumption of the inference program using $\verb|PowerTOP|$ \cite{powertop}. On the GPU platform, we measure power consumption using $\verb|nvidia-smi|$ \cite{nvidia-smi} command tool. For the FPGA board (ZCU 104), we use an external power meter to measure {its} power consumption. The reported numbers in Table \ref{tab:energy-comparison} are the average power consumption during inference. The results show that our FPGA accelerator is $62\times$, $39\times$ more energy efficient than CPU and GPU platform, respectively.


\section{Conclusion}

In this paper, we propose a novel model-architecture co-design for SAR ATR on FPGA. The proposed lightweight GNN model  achieves similar accuracy with state-of-the-art {models} with only $1/3258$ computation complexity and $1/83$ model size. The proposed accelerator on an embedded FPGA platform has lower latency than the state-of-the-art CPU/GPU with significant less energy consumption. 

\section*{Acknowledgment}
This work is supported by the National Science Foundation (NSF) under grants OAC-1911229, CNS-2009057, and in part by DEVCOM Army Research Lab (ARL) under ARL-USC collaborative grant DIRA-ECI:DEC21-CI-037. The author Bingyi Zhang is supported by the Summer Research Program from the Army Research Lab West (ARL West).

\bibliographystyle{IEEEtran}
\bibliography{main_full}

\end{document}